\documentclass[aps,twocolumn,showpacs,amssymb,amsmath]{revtex4}
\usepackage{graphicx}
\usepackage{epsfig}
\usepackage{bm}

\begin{document}

\title{Peak Effect in Superconductors: Absence of Phase Transition and Possibility of
Jamming in Vortex Matter\footnote{This work was carried out during the authors' stay
in the Department of Physics, University of California, Davis, 95616.}}

\author{Mahesh Chandran\footnote{Present Affiliation: Materials Research Laboratory, John
F Welch Technology Centre, GE India Technology Centre, Phase 2, Hoodi Village,
Whitefield Road, Bangalore, 560066; Email address: mahesh.chandran@ge.com}}

\affiliation{ }

\date{\today}

\begin{abstract}
                                                                                           
The magnetic field $B$ dependence of the critical current $I_c$ for the vortex phase of a
disordered superconductor is studied numerically at zero temperature. The $I_{c}(B)$
increases rapidly near the upper critical field $B_{c2}$ similar to the peak effect (PE)
phenomenon observed in many superconductors. The real space configuration across the PE
changes continuously from a partially ordered domain (polycrystalline) state into an
amorphous state. The topological defect density $n_{d}(B)\sim e^{\alpha B^{k}}$ with $k>1$
for $B\geq 0.4B_{c2}$. There is no evidence of a phase transition in the vicinity of the
PE suggesting that an order-disorder transition is not essential for the occurrence of the
PE phenomenon. An alternative view is presented wherein the vortex system with high
dislocation density undergoes jamming at the onset of the PE.
                                                                                           
\end{abstract}
                                                                                           

\maketitle
                                                                                           
\section{Introduction}

The critical current density $J_{c}$ for a type-II superconductor decreases monotonically
as a function of the magnetic field $B$ or the temperature $T$ over most part of the
$B$-$T$ phase diagram. But quite often, a peak in the $J_{c}(B)$ (or $J_{c}(T)$) is
observed close to the upper critical field $B_{c2}(T)$ (or critical temperature
$T_{c}(B)$). This phenomenon is known as the peak effect (PE) and was first studied in
1961. Extensive work since then have shown that PE exist in wide class of materials which
includes low-$T_c$ (Nb and its alloys\cite{berlincourt}, V$_3$Si\cite{isino},
NbSe$_2$\cite{higgins}), high-$T_c$ (YBa$_2$Cu$_3$O$_7$\cite{danna}), heavy fermion
(UPt$_3$\cite{tenya}, UPd$_2$Al$_3$\cite{ishiguro}), and other exotic
(Sr$_2$RuO$_4$\cite{tamegai}) superconductors, including the recently discovered compound
MgB$_2$\cite{pissas}. Apart from single crystals, the phenomenon have been observed in
polycrystalline and thin amorphous films\cite{wordenweber1}. The PE is found to be more
pronounced in the weak pinning limit and progressively diminishes with increasing pinning
strength\cite{banerjee1,andrei}. An important characteristic of the PE phenomenon is the
strong memory and history dependence of the various properties, and have been observed in
static and dynamic experiments\cite{good,xiao1,henderson1,valenzuela,chaddah,banerjee2}.

The explanation for the PE is based on the softening of the vortex lattice with increasing
$B$, first suggested by Pippard\cite{pippard}. If the vortex lattice rigidity falls more
rapidly than the elementary vortex pinning strength $f_{p}$ as $B\rightarrow B_{c2}$, the
softening would allow vortices to conform to the random pinning potential better compared to
a stiff vortex lattice. This idea was further developed and quantified in the collective
pinning theory by Larkin and Ovchinnikov\cite{larkin} (LO). In the LO theory, the vortex
lattice is broken into coherently pinned regions of volume $V_{c}=R_{c}^{2}L_{c}$ within
which the pinning forces act randomly. $R_{c}$ and $L_{c}$ are the transverse and the
longitudinal correlation lengths, respectively, and are related to the tilt modulus $C_{44}$
and the shear modulus $C_{66}$ of the vortex lattice. The balancing of elastic energy and
the pinning energy within $V_{c}$ gives the critical current density $J_{c}\propto
f_{p}(n_{p}/V_{c})^{1/2}$. LO suggested that in $3D$ the spatial dispersion of $C_{44}$ can
lead to an exponential decrease in $V_c$ near $B_{c2}$ giving rise to the PE phenomenon. 
The PE is not expected in the $2D$ since the only relevant elastic modulus $C_{66}$ is 
weakly dispersive.

Experimentally, the PE have been observed in thin amorphous films (samples with thickness $d
< L_{c}$). This was interpreted within the collective pinning theory as due to dimensional
crossover from $2D$ to $3D$ in the pinning characteristic\cite{kes}. Recently, the PE have
been observed in ultra thin Bi films of $d= 26 A^{\circ}$\cite{samband} which is difficult
to account for within the dimensional crossover scenario. Moreover, neutron scattering
experiment on Nb crystal shows that $L_{c}$ actually increases with $B$, and the decrease in
$L_{c}$ in the PE region is not appreciable\cite{gammel}. Also, the LO theory cannot 
account for the PE in $J_{c}(T)$\cite{ling}.

In the LO picture, the pinning energy overcomes the elastic energy beyond $R_{c}$, and
hence, the long range order is not expected to survive in the vortex lattice. In high-$T_c$
superconductors, topologically ordered vortex lattice on a length scale $r\gg R_{c}$ was
observed\cite{grier,forgan}. Detailed calculations showed that in the presence of weak
point impurities a quasi-long range ordered vortex lattice survives in $3D$ with $C(r)\sim
r^{-\eta}$, where $\eta$ is a non-universal exponent\cite{giamarchi1,nattermann}. Such a
phase has been termed as the Bragg glass (BG). The BG undergoes a first-order melting
transition into a vortex liquid (VL) phase on increasing $T$. Also, the BG undergoes a weak
first-order transition into the vortex glass (VG) phase on increasing $B$\cite{giamarchi2}.
In the VG phase, the plastic deformation caused by the topological defects leads to strong
pinning, and hence higher $J_{c}$ compared to the BG phase. The PE have thus been
interpreted as a consequence of an order-disorder transition\cite{paltiel1,ravi}.  
Furthermore, the origin of anomalous dynamics near the PE is attributed to the coexistence
of BG and VG phases in the vicinity of the transition\cite{marchevsky,paltiel2}.

Interpretation of the PE phenomenon as an order-disorder transition has limitations.  
Firstly, the BG phase is not expected in $2D$\cite{zeng}, and hence order-disorder
transition cannot account for the PE in amorphous films. Secondly, the BG phase have been
shown theoretically to exist in the presence of weak point impurities. It is not clear if
the BG would survive in polycrystalline samples in which the PE has been well documented.
Recent Bitter decoration of NbSe$_2$ across the PE shows no distinct ordered and disordered
phases\cite{fasano}. Also, the width of the field range in which the ordered and the
disordered phases coexist have been shown to scale with the sample size, which is an
evidence against the phase transition interpretation of the PE\cite{schleser}.

Attempts have been made to understand the PE phenomenon using numerical simulation. Cha and
Fertig\cite{cha} used an attractive term in the inter-vortex potential to tune the $C_{66}$
in $2D$. A small peak in the total depinning force was observed which was associated with
the transition from the elastic to plastic depinning. But as emphasized in
Ref.\cite{charles1}, the peak in total depinning force as a function of $C_{66}$ does not
imply a peak in $J_{c}(B)$. Later simulation in $3D$ showed an insignificantly small peak
near the BG to VG transition\cite{anna}. The PE was also studied in a model of layered
superconductors where the peak in $J_c$ was induced through the decoupling
transition\cite{olson}.

In this paper, the behaviour of the critical current $I_{c}(B)$ for a $2D$ disordered
vortex matter is studied using numerical simulation. The effect of finite vortex core is
included in the inter-vortex interaction and is shown to be essential for the realistic
description of the system close to $B_{c2}$. The $I_{c}(B)$ decreases monotonically
in the intermediate field range and rises rapidly close to $B_{c2}$ similar to 
the PE phenomenon observed in real systems. The vortex configuration shows no
order-disorder transition but a continuous transformation from a polycrystalline domain
state below the PE to a liquid-like amorphous state in the PE region. An alternative
scenario for the PE is discussed where the vortex system undergoes jamming due to 
increased dislocation density. An empirical relation between the $I_{c}(B)$ and the real 
space defect density is also obtained from the simulation which captures the essential 
physics governing the behaviour of $I_{c}(B)$ for a disordered vortex matter.

The paper is organized as follows: in section 2, the simulation details are presented,
followed by results in section 3. The results are discussed in Section 4, and the
conclusions are summarized in Section 5.

\section{Simulation method}

Consider a $2D$ cross-section of a bulk type-II superconductor perpendicular to the
magnetic field ${\bf B}=B\hat{\bf z}$. We model the vortices as classical particles 
interacting via two-body potential, and governed by an overdamped equation of motion
\begin{eqnarray} 
\eta\frac{d{\bf r}_{i}}{dt} = -\sum_{j\neq i} \nabla
U_{C}^{v}({\bf r}_{i}-{\bf r}_{j}) - \sum_{k} \nabla U^{p}({\bf r}_{i}-{\bf R}_{k}) + {\bf 
F}_{ext},
\end{eqnarray} 
where $\eta$ is the flux-flow viscosity. The first term on the left hand side represents
the inter-vortex interaction, and the second term is the attractive interaction between the
vortex and the quenched impurities. ${\bf F}_{ext}$ is the Lorentz force on the vortex due
to an applied current. Each of the three terms are further discussed in detail below.

The inter-vortex potential $U_{C}^{v}(r)=\frac{\phi_{0}^{2}}{8\pi^{2}\lambda^{2}}
K_{0}(\tilde{r}/\lambda)$, where $K_0$ is the modified Bessel function, and $\tilde{r} =
(r^{2}+2\xi^{2})^{1/2}$. $\phi_{0}$ is the flux quantum, and $\lambda$ and $\xi$ are the
penetration depth and the coherence length of the superconductor respectively. This form
of the interaction was derived by Clem from the Ginzbug-Landau (GL) equation using
variational method\cite{clem,brandt}. Conventionally, the vortex dynamics have been
studied using the potential $U_{L}^{v}(r)=\frac{\phi_{0}^{2}}{8\pi^{2}\lambda^{2}}
K_{0}(r/\lambda)$ which is derived from the London's equation. For small $r$, the
$K_{0}(r)\propto -\ln(r)$, leading to an unphysical singularity in $U_{L}^{v}(r)$ as
$r\rightarrow 0$. The potential $U_{L}^{v}(r)$ is thus inadequate for describing
correctly the vortex core region which limits its applicability to small $B$. On the
other hand, Clem's potential $U_{C}^{v}(r)$ is well behaved as $r\rightarrow 0$ and gives
realistic form of the magnetic flux density ${\bf B}(r)$ and the supercurrent density
${\bf J}(r)$ near the vortex core region. The correct description of the vortex core
region is essential to extend the simulation to high magnetic fields where the cores of
the neighboring vortices tend to overlap.

An important consequence of using Clems's potential is that the vortices are not point
particles but have a finite radius $\xi$. This is an improvement over the previous
simulations which treated vortices as point particles interacting via London's potential
$U_{L}^{v}(r)$. Note that the London's model of vortices interacting via two-body potential
is strictly valid for $B<0.25 B_{c2}$\cite{brandt}. Even though Clem's potential overcomes
some of the limitations of the London's potential at high fields, it still provides only a
qualitative behaviour of the vortex system. A more realistic approach for understanding the
vortex dynamics close to $B_{c2}$ is through the time dependent GL theory (TDGL).
Unfortunately, the computation cost for TDGL is prohibitive even for a small sample 
size\cite{crabtree} which restricts its utility.

The second term in Eq.(1) is the attractive force between the vortex and the quenched
impurities which act as pinning centers. The force is derived from the parabolic potential,
$U^{p}(r)=U_{0}(\frac{r^{2}}{r_{p}^{2}}-1)$ for $r<r_{p}$, and 0 otherwise. The impurities are
randomly placed at positions ${\bf R}_{k}$ in the simulation box. The third term is the
Lorentz force ${\bf F}_{ext}=\frac{1}{c}{\bf J}\times \phi_{0}\hat{\bf z}$ where ${\bf J}$ is
the the current density. A uniform current density along the $y$-direction is assumed such
that all vortices experience the same force $F_{ext}^{x}$ along the $x$-direction. For the
simulation purpose, the length is defined in units of $\lambda(B=0)=\lambda_{0}$.  The current
density $J$ and the velocity of the vortex $v_{x}$ are in the units of $cf_{0}/\phi_{0}$ and
$f_{0}/\eta$, respectively, where $f_{0}=\frac{\phi_{0}^{2}}{8\pi^{2}\lambda_{0}^{3}}$. From
the relation ${\bf E}={\bf v}\times{\bf B}$, the $v_{x}\propto E_{y}\propto V$ where $V$ is
the voltage generated in the direction of the current $I\propto J_{y}\equiv F_{ext}^{x}$.
Hence, the $v_{x}(F_{ext}^{x})$ behavior represents the $V(I)$ characteristic (or equivalently
$E(J)$ curve) for a superconductor.

The dimensionless magnetic field is defined as $b=B/B_{c2}$, where the upper critical field
$B_{c2}=\frac{\phi_{0}}{2\pi\xi_{0}^{2}}$ and $\xi_{0}=\xi(B=0)$. The $b$ is calculated from
the lattice constant
$\frac{a_{0}}{\lambda_0}=(\frac{4\pi}{\sqrt{3}})^{\frac{1}{2}}(\frac{1}{\kappa^{2}b})^{\frac{1}{2}}$.
The GL parameter $\kappa=\frac{\lambda}{\xi}$ is an input to the simulation. In the GL theory,
the length scale $\lambda$ and $\xi$ increases with $b$ and diverges as $b\rightarrow 1$. The
renormalization of $\lambda$ and $\xi$ with increasing $b$ is included in the simulation
through the relation $\lambda(b)=f(b)\lambda_{0}$ and $\xi(b)=f(b)\xi_{0}$, where
$f(b)=1/\sqrt{(1-b^{2})}$. This form of $f(b)$ is similar to the $T$-dependence of $\xi$ in
the GL theory\cite{tinkham} with $T/T_{c}$ replaced by $(B/B_{c2})^{2}$, and have been used
previously in Ref.\cite{ryu}. The same form of $f(b)$ is chosen for $\lambda$ and $\xi$ in
order to keep $\kappa$ independent of $B$.

The simulation was carried out by numerically integrating the equation of motion using the
fourth order predictor-corrector scheme. Parallel algorithms were implemented details of 
which can be found in Ref.\cite{chandran1}. A perfect vortex lattice
driven by a large current $I\gg I_{c}$ was used as the starting condition for the
simulation. The $I$ is then reduced to 0 in small steps and the average voltage $V$ was
calculated in the steady state at each step to obtain the $V(I)$ curve. In the dimensionless
units defined above, the $V=I$ in the asymptotic limit $I\rightarrow\infty$. The $I_{c}$ is
defined as the current at which the $V\lesssim 10^{-5}$ \cite{comment-ic}. The procedure
used for obtaining the $V(I)$ curve was motivated by experiments where a stable vortex
configuration is observed if the system is brought to rest after driven with $I\gg
I_{c}$\cite{henderson2}. Also, the $I_{c}(b)$ from such a method shows the largest peak
compared to that obtained from the field-cooling experiments.

The real space configuration of vortices was characterized by locating the topological
defects in the system using Delaunay triangulation. For the triangular vortex lattice, a
topological defect is a vortex with coordination number other than 6. The defect
density $n_{d}/\lambda_{0}^{2}$ is defined as the number of defects per unit area of the
simulation box. Most of the defects form dislocations which are bound pair of 
disclinations (vortices with coordination 5 and 7). The fraction of free disclination is small
over a wide range of fields, and hence $n_{d}\approx 2n_{disl}$ where $n_{disl}$ is the
dislocation density. Along with the $V(I)$ curve, the behaviour of $n_{d}(I)$ was also
determined for each value of $b$. The $n_{d}$ discussed below represents the defect
density at $I=0$, unless specified otherwise.

The parameters used in the simulation are $\kappa=10$ and $\lambda(B=0)=1000\AA$. These
values are typical of low-$T_{c}$ superconductors, particularly NbSe$_2$. The periodic
boundary conditions were imposed in both directions. The magnetic field $b$ was varied by
changing the size of the simulation box, keeping the number of vortices $N_{v}$ fixed
(for smaller system size, $N_{v}$ is also allowed to vary between 800-1200). The results
presented below are for $N_{v}=4096$. The prefactor $U_{0}$ of the pinning potential
$U^{p}(r)$ is distributed randomly between $\Delta \pm 0.01$ where $\Delta = \langle
U_{0}\rangle$. The simulation is restricted to point impurities with the range of the
pinning potential $r_{p}=\xi_{0}$. The pin density $n_{p}=2.315/\lambda_{0}^{2}$. The
dilute limit of $n_p$ is chosen keeping in mind that the PE is observed close to $B_{c2}$
where the concentration of effective pinning centre is expected to be very small.

\begin{figure}[th]
\centerline{\psfig{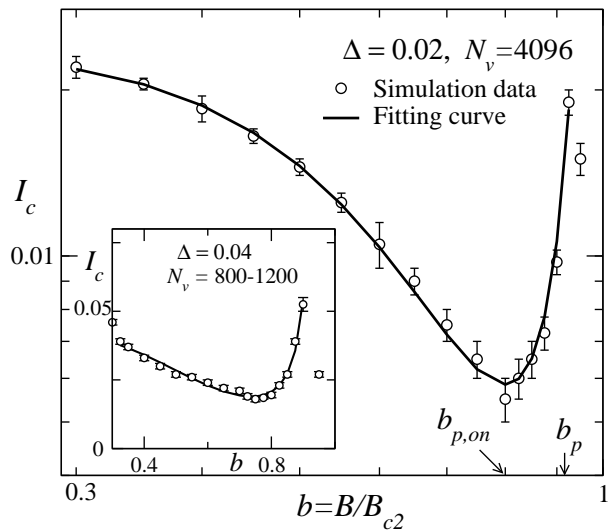}}
\vspace*{8pt}
\caption{The critical current $I_{c}(b)$ for $\Delta=0.02$ and $N_{v}=4096$. The $I_{c}(b)$ 
for $\Delta=0.04$ is shown in the inset. For this plot, the $N_{v}$ was also varied between 
800 and 1200 along with the simulation box size. The thick line represents the fit to the 
simulation data (shown by open symbols).}
\end{figure}

\section{Results}

\subsection{Critical current $I_{c}(b)$} 

Figure 1 shows the $I_{c}(b)$ for $\Delta=0.02$. For $b<b_{p,on}=0.8$, the $I_c$ decreases
monotonically. At $b_{p,on}$, the $I_{c}(b)$ turns around and rises rapidly till $b=0.925$.
The rapid increases in $I_{c}$ above $b_{p,on}$ signals the appearance of the PE. The inset
shows the PE for $\Delta=0.04$ for which a smaller system size was used. The $b_{p,on}$
decreases from a value of 0.9 for $\Delta=0.01$ to 0.75 for $\Delta=0.04$. The proximity of
the PE to the upper critical field $B_{c2}$ is in good agreement with the experiments.

For $b>b_{p}=0.925$, the $I_{c}$ decreases. For $b\geq 0.95$ two vortices occasionally come
together to form an effective vortex with charge $2\phi_{0}$. Overlapping of the two vortex
cores is strongly influenced by the impurities. This was confirmed by simulating without the
pinning centres where no overlapping of the vortices was found up to $b=0.98$. The effective
interaction between a $2\phi_{0}$ vortex and the $\phi_0$ vortex is greater than the
interaction between $\phi_0$ vortices. This enhances the local rigidity of the vortex
system, and possibly decrease the $I_{c}$. It is worth mentioning that a short range
attractive interaction exist between the vortex cores\cite{brandt} which is not included in
the simulation. The important point is that the overlapping of vortex cores is observed for
$b\geq 0.95$, and hence does not influence the increasing branch of the PE. It is possible
that the $I_{c}(b)$ above $b_{p}$ is influenced by the overlapping of vortices, a issue
which can be addressed only by solving TDGL equations in the presence of impurities.

Attempts were made to fit the $I_{c}(b)$ data to a single function across the PE. A 
reasonably good fit was obtained using the expression
\begin{equation}
I_{c}(b)=I_{c0}\exp(-\beta b^{k})\frac{1}{(1-b^{2})^{4}},
\end{equation}
with $k\approx 2.6$ and $\beta\approx 9.75$ for $\Delta=0.02$. For $\Delta=0.04$, $k\approx
2.4$ and $\beta\approx 8.3$. The above function was motivated by the behaviour of the
$n_{d}(b)$ which follows the same exponential form, whereas the dependence on
$\frac{1}{(1-b^{2})}$ reflects the behaviour of $\lambda$ and $\xi$ as $b\rightarrow 1$.
The renormalization of the length scales $\lambda$ and $\xi$ leads to weak inter-vortex
interaction, and explains why the PE in real system occurs close to the $B_{c2}(T)$ line.
As shown below, the bare length scale $\lambda_{0}$ and $\xi_{0}$ leads to a stiff vortex
lattice for $b>0.6$ and consequently the PE is absent.

\begin{figure}[th]
\centerline{\psfig{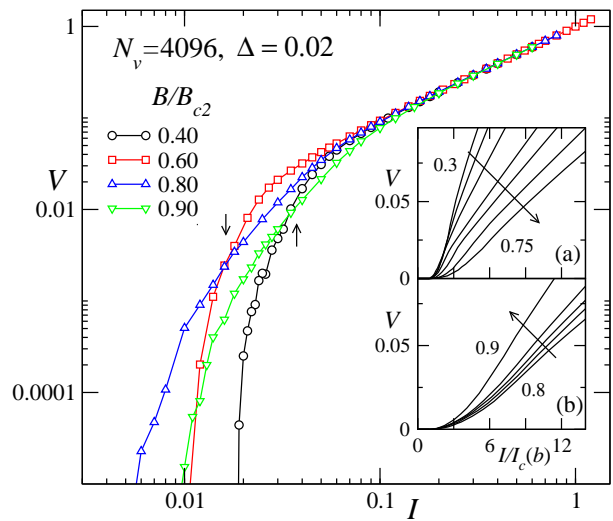}}
\vspace*{8pt}
\caption{The $V(I)$ curve across the PE. Arrows point to the crossing of $V(I)$ curves. 
Inset: the $V(I/I_{c}(b))$ curves for (a) $b<b_{p,on}$, and for (b) $b\geq b_{p,on}$. The 
arrows represent the direction of increasing $b$. The magnetic fields in (a) are 0.3, 0.4, 
0.5, 0.6, 0.65, 0.7 and 0.75. In (b), the $b$ increases from 0.8 to 0.9 in steps of 0.025.}
\end{figure} 

\subsection{$V(I)$ characteristics across the Peak Effect}

Figure 2 shows the $V(I)$ curves across the PE. The curves for high-$b$ crosses the
curves for low-$b$ in the sub-ohmic region. The crossing can be understood from the
dynamics close to $I_{c}$. For $I\gtrsim I_{c}$, the dynamics is heterogeneous with few
vortices moving in channels in a background of defective vortex
configuration\cite{jensen,charles2,dominguez}. The width of the channels $\sim a_{0}$ and
display large transverse wandering relative to the direction of the Lorentz force. Since
not all vortices are active, the voltage response is sub-ohmic. The number of active
channels increases with $I$, and for $I=I_{p}$ all vortices move in channels which are
ordered transverse to the flow direction. The time-averaged defect density shows an
abrupt drop at $I_{p}$, and represents annealing of the plastic deformation (some
dislocations with Burger vector parallel to the Lorentz force can exist for $I>I_p$). For
$I>I_{p}$, the response is ohmic and the resistance approaches the free flux-flow value.
The sub-ohmic response window $\delta I_{dis}=I_{p}-I_{c}$ thus quantify the extent of
plastic deformation of the vortex system. As shown in Ref.\cite{chandran2}, $\delta
I_{dis}(b)$ increases rapidly in the PE region implying that larger currents are required
to anneal the defects. This is reflected in the slow growth of $V(I)$ curve for high
fields, which consequently cross the curves for low fields as shown in Fig. 2.

The inset in Fig. 2 shows the change in $V(I)$ characteristics as $b$ is varied across
the PE. For each curve, the $I$ is scaled by the respective $I_{c}(b)$. Far below the PE,
the curves are convex for $I\gtrsim I_{c}$. As $b$ approaches $b_{p,on}$, the curvature
changes to concave. The concave part of the $V(I)$ curve indicates a slow growth of the
response due to increased plasticity of the system. Interestingly, for $I/I_{c}(b)>1$, 
the $V(b)$ reflects the non-monotonic behaviour of $I_{c}(b)$. Thus, the PE is 
characterized by static as well as dynamic changes in the system, and agrees with the 
conclusion from the experiments\cite{shobo}.

\begin{figure}
\centerline{\psfig{file=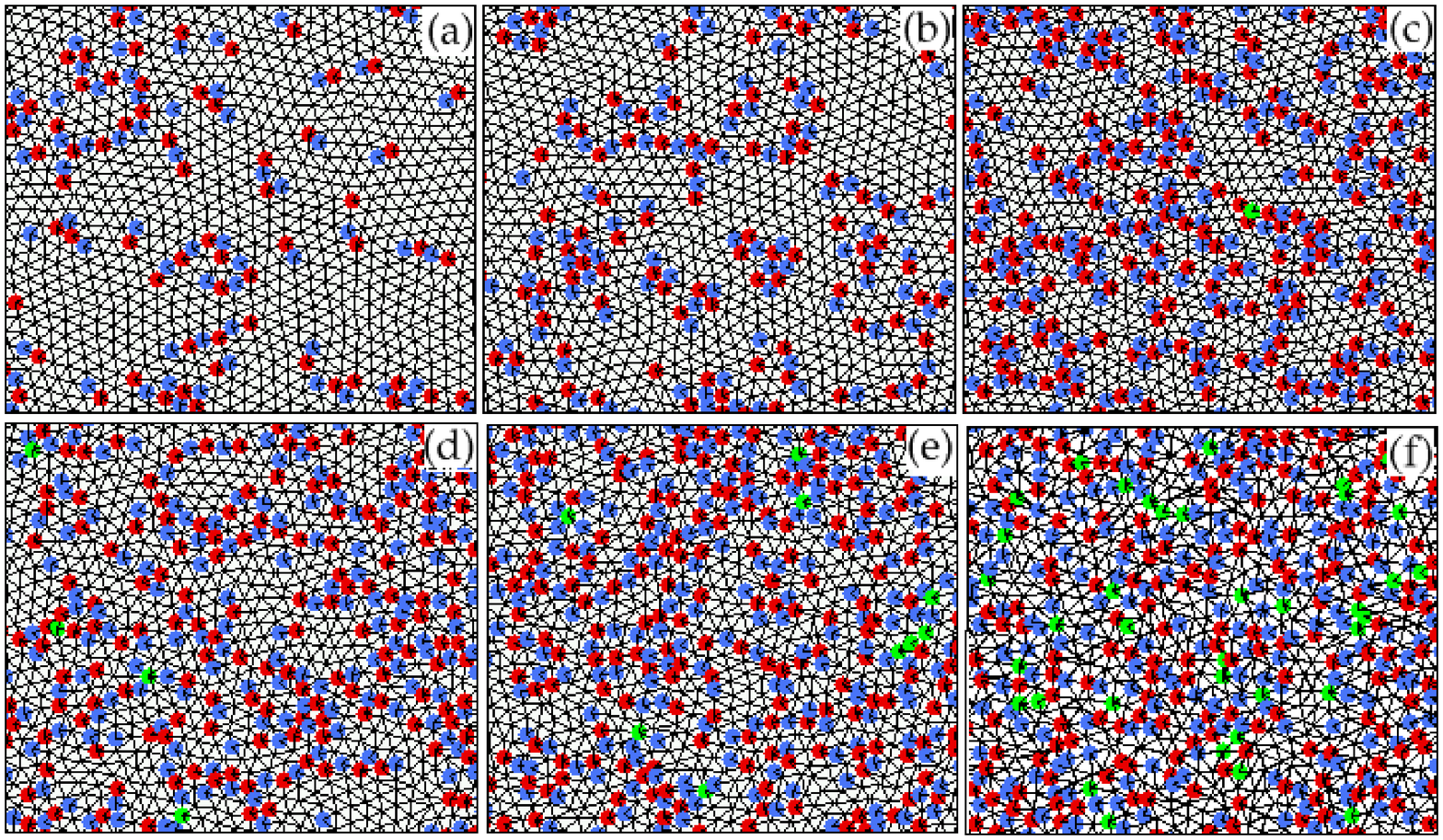,width=8.5cm}}
\vspace*{8pt}
\caption{The Delaunay triangulation of the real space configuration across $b_{p,on}$.
The $b$=0.70 (a), 0.75 (b), 0.80 (c), 0.85 (d), 0.875 (d), and 0.90 (f). The red and the 
blue dots represents vortices with 7 and 5 neighbors, respectively. The vortices with 
4 and 8 neighbors are denoted by the green dot. The pinning strength $\Delta=0.02$ 
and $N_{v}=4096$. Only a small region of the simulation box is shown for clarity.}
\end{figure} 

\subsection{Real space configuration}

The simulation allows the changes in the real space configuration to be characterized
precisely, thus allowing theories based on configurational changes to be tested. Fig. 3
shows the real space images at $I=0$ in a region of the simulation box as $b$ is changed
across $b_{p,on}$. The defects are marked by filled circles. A topologically ordered vortex
lattice is not observed for any value of $b$ consistent with the theory for $2D$
system\cite{giamarchi1}. But partial order can still be seen in large domains for $b$
between $0.3-0.6$. The dislocations are arranged in string-like structures and forms the
grain boundary (domain walls). There are no free dislocations inside the domain, and free
disclinations are absent below $b_{p,on}$. The configuration in the intermediate fields is
reminiscent of a polycrystalline solid. With increasing $b$, the domain size decreases but
the increase in the defect density occurs {\em within} the grain boundaries. Some of the
domains can also be seen for $b=0.70$ and 0.75. The average size of the domains decreases
from $\approx 6a_{0}$ below the PE to $\approx 3a_0$ on the increasing branch of
$I_{c}(b)$. At $b_p$, the configuration is amorphous (Fig. 3(d)) with defects forming a
dense homogeneous network. It is not possible to isolate individual dislocations and the
configuration is that of a frozen liquid\cite{comment-ndef}.

The continuous transformation of a less disordered state below $b_{p,on}$ into an
amorphous state at $b_p$ precludes an order-disorder (or BG to VG) transition underlying
the PE. The PE is thought to be due to such a phase transition in the vortex
system\cite{paltiel1}. The real space images shown in Fig. 3 is strikingly similar to the
recent Bitter decoration of NbSe$_2$ in the PE region\cite{fasano}, which lends support
to the idea that a phase transition is not necessary for the PE to occur. The amorphous
state at $b_{p}$ is also consistent with the neutron scattering experiment\cite{gammel}
and the STM imaging of the vortices\cite{troy} in the PE region.

\begin{figure}[th]
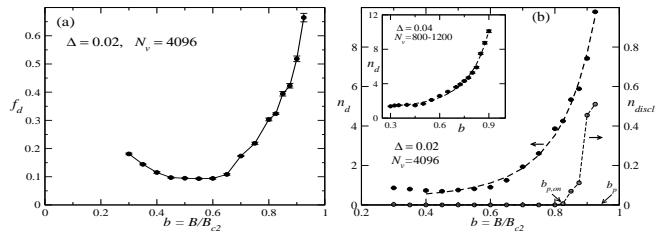

\centerline{\psfig{file=fig4a.eps,height=3cm,width=4.25cm}
\psfig{file=fig4b.eps,height=3cm,width=4.25cm}}
\vspace*{8pt}
\caption{(a) The defect fraction $f_{d}(b)$ at $I=0$. (b) The defect density $n_{d}(b)$ 
at $I=0$. The $n_{discl}=|n_{d,7}-n_{d,5}|$ is the approximate density of free 
disclinations, where $n_{d,7}$ and $n_{d,5}$ are the density of vortices with 
coordination number 7 and 5, respectively. Inset: the $n_{d}(b)$ plot for $\Delta=0.04$. 
The thick dashed line is a fit to $n_{d}(b)=n_{d0}\exp(\alpha b^{k})$.}
\end{figure}

\subsection{Defect density $n_{d}(b)$}

Figure 4(a) shows the defect fraction $f_{d}(b)$, defined as the ratio of number of
defects to number of vortices, for $\Delta=0.02$. As mentioned earlier, reducing $I$ to
zero from $I\gg I_{c}$ generates a configuration which is stable and does not show
hysteresis\cite{henderson2}. The $f_{d}(I=0)$ can thus be considered as the equilibrium
defects which determines the behaviour of stable $I_{c}(b)$. The $f_{d}$ is typically
greater than 0.35 for $b<0.1$. Similar value for $f_{d}$ is reached at the onset of PE.
The intermediate polycrystalline state is sandwiched between $0.3\lesssim b\lesssim 0.7$, 
with $f_{d}$ less than 0.2.

Interestingly, if the defect density $n_{d}(b)$ rather than $f_{d}(b)$ is plotted, a good 
fit can be obtained for $b\gtrsim 0.4$ using the expression
\begin{equation} 
n_{d}(b)=n_{d0}\exp(\alpha b^{k}).
\end{equation} 
This is shown in Fig.4(b) for $\Delta=0.02$ and $\Delta=0.04$ (inset). The exponent 
$k\approx 2.6$ and $2.4$ for $\Delta=0.02$ and $0.04$, respectively. Surprisingly, both 
$n_{d}(b)$ and $I_{c}(b)$ can be fit using the same value of $k$. This allows one to 
write
\begin{equation} 
I_{c}(b)=\frac{I_{c1}}{[n_{d}(b)]^{r}}\frac{1}{(1-b^{2})^{4}},
\end{equation} 
where $r\approx 2.5$ and $2.7$ for $\Delta=0.02$ and $0.04$, respectively. Figure 4(b) also
shows the density of free disclinations which can be approximately defined as
$n_{discl}=|n_{d,7}-n_{d,5}|$, where $n_{d,7}$ and $n_{d,5}$ are the density of vortices
with coordination number 7 and 5, respectively. The $n_{discl}$ is 0 for $b < b_{p,on}$, and
rises rapidly in the PE region.

From the analysis of the $n_{d}(b)$ and the $I_{c}(b)$ data, it is possible to conclude
that $I_{c}(b)\propto [n_{d}(b)]^{-r}$, where the effect of the pinning potential is
implicit through the defect density $n_{d}(b)$ and the exponent $r$. This empirical
relation between $n_{d}(b)$ and $I_{c}(b)$ captures the essential physics governing the
behaviour of the critical current in a disordered vortex system with point impurities.
Physically, the increase in $n_{d}$ with $b$ enables plastic shearing to be
initiated at lower currents, which explains the monotonic decrease in $I_{c}$ below the
PE. But this also leads to a paradoxical situation: since the $n_{d}(b)$ continues to
rise rapidly even in the PE region, the enhanced plasticity of the system should cause
the $I_{c}$ to decrease continuously till $B_{c2}$, contrary to the observation of PE
above $b_{p,on}$. The problem is compounded further since the real space images do not
suggest any noticeable configurational change across $b_{p,on}$ which can be associated
with sudden increase in pinning due to quenched impurities.

The above paradox can be resolved by conjecturing that the vortex system undergoes
jamming at the onset of the PE due to increase in dislocation density. The jamming
envisaged here is similar to the jamming in granular materials where the system develops
a finite yield stress at a threshold packing fraction\cite{hern}. The jamming have been 
also studied in dislocation network formed in a plastically deformed material\cite{miguel} 
which also shows complex spatio-temporal dynamics\cite{gdanna}. The jamming in the
vortex system could occur due to disordered packing of vortices which imposes kinematic
constraint for the motion of vortices. From Fig. 4(a), the defect fraction $f_{d}$
becomes 0.30 at $b_{p,on}$, which could possibly be the threshold value for the jamming
to occur in the vortex system. As mentioned earlier, the dynamics close to the $I_{c}$ is
governed by few active channels. The effect of disordered packing of vortices is to
increase the effective energy barrier $\Delta E_{T}$ for the transverse mobility of the
channel, thus requiring larger current to initiate channel dynamics. Future work should
be able to obtain the $B$-dependence of $\Delta E_{T}$, which could provide evidence for
jamming in the PE region.

\begin{figure}[th]
\centerline{\psfig{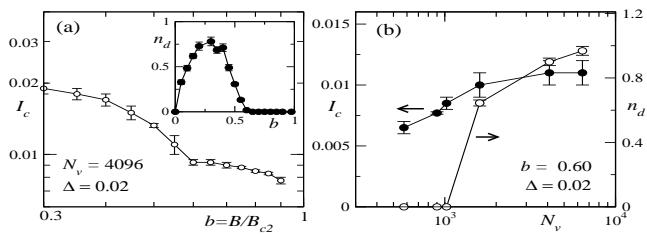}}
\vspace*{8pt}
\caption{(a) The $I_{c}(b)$ and $n_{d}(b)$ (inset) without the field dependence of
the length scales $\lambda$ and $\xi$. The $\Delta=0.02$ and $N_{v}=4096$. (b) The $I_{c}$ 
and $n_{d}$ as a function of $N_{v}$ for $b=0.6$ and $\Delta=0.02$ }
\end{figure}

\subsection{Field independent $\lambda$ and $\xi$}

The $I_{c}(b)$ and the $n_{d}(b)$ obtained using $\lambda(b)=\lambda_{0}$ and
$\xi(b)=\xi_{0}$ are shown in Fig. 5(a). The $\lambda_0$ and $\xi_0$ are the values at
$B=0$. For $b<0.6$, the $I_{c}(b)$ does not show any appreciable deviation from that of
Fig. 1. At $b\approx 0.6$, the defect density $n_{d}$ becomes zero, and the system
remains topologically ordered till $B_{c2}$. As a consequence, the PE is absent at high
fields and is consistent with the idea that the PE cannot occur in rigid vortex lattice.

\subsection{Finite size effect}

The finite size effect on $n_{d}$ and $I_{c}$ was determined by increasing $N_{v}$ for
$b=0.6$. The total number of defects shows a minimum around $b=0.60$, or equivalently,
the average size of the ordered region (domain) is maximum for this field. Hence, any
relevant length scale determining the system behaviour would be largest for $b=0.6$.
Figure 5(b) shows $I_{c}$ and $n_{d}$ as a function of the system size ($\propto N_{v}$).
The maximum $N_{v}$ which could be simulated was 6400. For $N_{v}\gtrsim 4000$, the
$I_{c}$ tends to become independent of the system size though $n_{d}$ continues to 
increase, suggesting that the two quantities are determined by different length scales.
The finite size calculation also shows that the system size used is sufficient to obtain
macroscopic behaviour of the $I_{c}(b)$, and demonstrates the necessity of going to 
larger system size for doing realistic simulation of the vortex dynamics.

\section{Discussions}

The results presented above clearly demonstrates that a phase transition is not essential
for the PE to occur in superconductors. This rules out any generic explanation based on
order-disorder (or Bragg glass to vortex glass) transition, and resolves the difficulty in
interpreting the PE in thin amorphous films. Also, the simulation results cannot be
explained within LO theory. The $R_{c}$ in the absence of dislocations in $2D$ was shown to
be $\sim 4-5 a_{0}$\cite{chandran3}, which is much smaller than the length scale
determining the behaviour of $I_{c}$. Further, the amorphous regime also exists for
$b\lesssim 0.2$ and occurs due to single vortex pinning. If the amorphous state in the PE
regime is attributed to the single vortex pinning, one expects similar dynamical behaviour
in the PE regime and in low fields. Contrarily, the anomalous dynamics is observed only in
the PE regime, thus indicating that one need to go beyond the pinning concept to describe
the vortex state in the PE regime.

As discussed in Section 3.4, an alternate scenario based on jamming can be invoked to
explain the increase in $I_{c}$ with increasing $n_{d}$. One of the important feature of a
jammed state is the athermal relaxation under applied stress. For example, logarithmic
relaxation is observed as a function of the number of ``taps'' applied to a a granular
heap\cite{knight}. Evidence for similar behaviour exists in vortex system in the PE region.
Henderson {\it et al.}\cite{henderson1} showed that the voltage response with
bi-directional current grows with the number of pulses before saturating, which is an
evidence for athermal behaviour. In an interesting recent experiment Ravikumar {\it et
al.}\cite{ravikumar} measured the magnetic relaxation in V$_3$Si in the PE region. The
authors found that the magnetic moment $M$ as a function of serial number of measurement
$n$ (rather than the real time) follows logarithmic behaviour. If the measurement $n$ is
assumed to be the number of ``taps'' applied to the system\cite{comment-ravi}, the athermal
relaxation observed is similar to that observed in Ref.\cite{knight}, indicating a common
underlying mechanism in the two systems. More experiments and simulations are necessary to 
probe the 
jamming scenario in the vortex system in the PE region.

\section{Conclusions}

The paper describes the results from detailed simulation of the magnetic field dependence of
$I_{c}$ for a superconductor at $T=0$. The $I_{c}(B)$ shows a large increase near the upper
critical field $B_{c2}$ in agreement with the experimental observation of the peak effect
phenomenon. There is no evidence of a phase transition in the peak effect region. Instead, a
continuous transformation of a less disordered polycrystalline state into an amorphous state
is observed implying that an ordered state is not necessary for the PE to occur. A jamming
scenario is presented which could explain the PE and can account for the anomalous dynamics
in the peak effect region.

\section*{Acknowledgments}

M.C. acknowledges useful discussions with A. K. Grover, E. Zeldov, S. Bhattacharaya, and
Ravikumar during various stages of the work, and thanks G. T. Zimanyi and R. T.
Scalettar for critical comments during the part of the work. The simulation was
performed at the Albuquerque High Performance Computing Center, University of New Mexico,
during the authors' stay at the University of California, Davis.


\begin{thebibliography}{0}
\bibitem{comment-ic} The $I_{c}$ can also be defined as the current at which the dynamic
resistance $\frac{dV}{dI}$ first becomes 0 as $I$ is reduced. The $I_{c}$ from this
definition does not differ more than $2\sigma$ from the criterion used, where $\sigma$ is
the step by which the current is changed in obtaining the $V(I)$ curve around $I_{c}$.
\bibitem{comment-ravi} In Ref.\cite{ravikumar}, the magnetization $M$ was measured in a
SQUID magnetometer where the sample undergoes excursion in a small inhomogeneous field.  
This generates a driving current near the surface of the sample. The Lorentz force due to 
this additional current can be thought as providing the ``tapping'' force for the vortex 
system.
\end{thebibliography}
\end{document}